\documentclass{cjaa}
\usepackage{graphicx,times}
\input{epsf.sty}
\input{psfig.sty}
\newcommand{\bsex}{\hbox{\tt SExtractor}}
\newcommand{\etal}{\hbox{et al.\,}}
\newcommand{\Kv}  {\hbox{$K_{\rm Vega}$}}
\newcommand{\sbzks}{\hbox{sBzKs}}

\newcommand{\zph}  {\hbox{$z_{\rm phot}$}}
\newcommand{\zsp}  {\hbox{$z_{\rm spec}$}}
\newcommand{\gsim}{\lower.5ex\hbox{$\; \buildrel > \over \sim \;$}}
\newcommand{\lsim}{\lower.5ex\hbox{$\; \buildrel < \over \sim \;$}}
\newcommand{\psim}{\lower.5ex\hbox{$\; \buildrel \propto \over \sim \;$}}
\def\h1{$h^{-1}$}

\begin{document}
\title{Star-Forming Galaxies at $z \sim 2$ in the Hubble
Ultra Deep Field\,$^*$ \footnotetext{$*$ Supported by the National
Natural Science Foundation of China.} }
\volnopage{Vol.\ 8 (2008), No.\ 1,~ 1--11}
   \setcounter{page}{1}
\author{Xu Kong
    \inst{1,2}
  \and Wei Zhang
    \inst{1,3}
  \and Min Wang
    \inst{1}
}
\institute{Center for Astrophysics, University of Science and
Technology of China, Hefei 230026; {\it xkong@ustc.edu.cn}\\
    \and
    Joint Institute for Galaxy and Cosmology (JOINGC) of SHAO and
USTC\\
    \and
    National Astronomical Observatories, Chinese Academy of 
    Sciences, Beijing 100012, China\\
\vs\no
   {\small Received 2007 April 12; accepted 2007 May 16}
}

\abstract{ Using a simple color selection based on $B$-, $z$- and
$K$-band photometry, $BzK= (z-K)_{\rm AB}-(B-z)_{\rm AB}>-0.2$, we
picked out 52 star-forming galaxies at $1.4\lsim z \lsim 2.5$ (\sbzks)
from a $K$-band selected sample ($\Kv<22.0$) in an area of $\sim 5.5$ 
arcmin$^2$ of the Hubble Ultra Deep Field (UDF). We develop a new 
photometric redshift method, and the error in our photometric 
redshifts is less than $0.02(1+z)$. From the photometric redshift
distribution, we find the $BzK$ color criterion can be used to
select star-forming galaxies at $1.4 \lsim z \lsim 2.5$ with
$\Kv<22.0$. Down to $\Kv<22.0$, the number counts of \sbzks\
increase linearly with the $K$ magnitude; the \sbzks\ are strongly 
clustered, and most of them have irregular morphologies on the ACS 
images.
They have a median reddening of $E(B-V)\sim0.28$, an average star 
formation rate of $\sim36\ M_{\odot}$~yr$^{-1}$ and a typical stellar 
mass of $\sim10^{10}M_\odot$. The UV criterion for the galaxies at 
$z\sim2$ can select most of the faint \sbzks\ in the UDF, but it does 
not work well for bright, massive, highly-reddened, actively 
star-forming galaxies.
\keywords{galaxies: evolution ---galaxies: high-redshift --- 
galaxies: photometry --- cosmology: observations}
}
\authorrunning{X. Kong et al.}
\titlerunning{sBzKs in the UDF}
\maketitle

\section{Introduction}\label{sec:int}

A number of observations suggest that the era of $z\sim2$ is
important in galaxy evolution for various reasons: the cosmic star
formation rate begins to drop at $z\sim1-2$ from a flat plateau
at higher redshifts; the morphological type mix of field galaxies
changes remarkably at $z\sim1-2$; the number density of QSOs has
a peak at $z\sim2$ (Dickinson \etal  2003; Fontana \etal  2003;
Steidel \etal  2004; Kong \etal  2006; Richards \etal  2006).
However, galaxies in the redshift interval $1.4 \lsim z \lsim 2.5$
have been called the ``redshift desert" by some researchers because, 
to date, we lack the tools to pick out dim galaxies in that range.

Recently, Steidel's group has extended the UV technique for 
selecting LBGs (Lyman Break Galaxies) to $z<3$ using a 
$U_{\rm n}GR_{\rm s}$ color-color
diagram which isolates star-forming galaxies at $z\sim2$ (Erb
\etal 2003; Steidel \etal 2004; Adelberger \etal 2004). However,
star-forming galaxies can be selected as LBGs only if they are UV
bright (i.e. actively star forming) and not heavily reddened by
dust. Using the highly complete spectroscopic redshift database of
the K20 survey, Daddi \etal (2004) introduced a new criterion,
$BzK=(z-K)_{\rm AB}-(B-z)_{\rm AB}>-0.2$, in the $B$-, $z$- and
$K$-band photometry, for obtaining a virtually complete sample of
star-forming galaxies in the redshift range $1.4\lsim z\lsim 2.5$
(for convenience, we use the term \sbzks\ for the galaxies
selected in this way). This criterion is reddening independent for
star-forming galaxies in the selected redshift range, thus can be
used to select the reddest dust-extinguished galaxies. This should
allow a relatively unbiased selection of star-forming galaxies at
$z\sim2$ within the magnitude limit of the sample studied.

Based on $BRIzJK$ photometry obtained by combining Subaru optical
and ESO near-IR data over two separate fields and using the $BzK$
selection technique, Kong \etal (2006) has obtained complete samples
of $\sim 500$ candidate \sbzks\ which were identified over an area
of $\sim 920$ arcmin$^2$ to $\Kv=19.2$, of which 320 arcmin$^2$ are
complete to $\Kv=20$. Using HST/ACS-$I$ (F814W),
Subaru-$BVg'i'r'z'$, CFHT-$ui$ and KPNO/CTIO-$Ks$ data, Renzini
\etal (2008) has selected $\sim 1300$ \sbzks\ over an area of $\sim
6600$ arcmin$^2$ to $\Kv=19.2$ in the COSMOS survey (Scoville 
\etal 2007). The surface density, number counts, clustering,
reddening, star formation rates (SFRs) and stellar masses ($M_*$) of
these {\it bright} \sbzks\ ($\Kv < 20$ or $\Kv < 19.2$) were
analyzed in these papers. To select {\it faint} \sbzks~and study
their properties, deep infrared images are required. The Hubble
Ultra Deep Field (UDF) provides us with the deepest view to date 
of the visible universe. 
In this paper, we will use the space-based and ground-based UDF data 
to select a faint \sbzks\ sample (down to $\Kv=22$) to  study the 
properties of faint \sbzks, and compare the differences between faint 
and bright \sbzks.

This paper is organized as follows. Section~2 describes the UDF
observations and the method for obtaining the photometric catalog.
Section~3 describes the calculation and calibration of photometric
redshifts. Section~4 presents the selection, morphologies, number
counts, clustering and physical properties of \sbzks. Section~5
compares the samples selected with the $BzK$ technique to those
UV-selected galaxies at $z\sim2$. Finally, a brief summary is
presented in Section~6. For the sake of comparison with previous
works, magnitudes and colors in both AB and Vega systems 
had to be used.

\section{UDF Overview}\label{sec:data}

The UDF (RA=$03^{\rm h}32^{\rm m}39^{\rm s}0$,
Dec=$-27\degr47\arcmin29\farcs1$, J2000) is located within one of
the best studied areas of the sky: the Chandra Deep Field South 
(CDF-S or GOODS-South region). With a total of 544 orbits, it is one 
of the largest time allocations with HST, and indeed the filter 
coverage, depth, and exquisite quality of the UDF ACS and NICMOS 
images provide an unprecedented data set for the study of galaxy 
evolution.  The field has been imaged by a large number of telescopes 
at a variety of wavelengths. In this paper, three sources of imaging 
data and the VLT/FORS2 spectroscopic data for this field are used.

\subsection{Observations}\label{sec:obs}

The UDF is a public $HST$ survey of a single Advanced Camera Survey
(ACS) wide field camera (WFC) field (11.5 arcmin$^{2}$)
in 4 broad-band filters: F435W ($B$), F606W ($V$), F775W ($i$) and
F850LP ($z$) (Beckwith \etal 2006). For our analysis we use the
reduced UDF data v1.0 made public by the Space Telescope Science
Institute on 09 March 2004. The 10-$\sigma$ limiting magnitudes at
$B$-band, $V$-band, $i$-band and $z$-band are 28.7, 29.0, 29.0 and
28.4, respectively, in an aperture of 0.2 arcsec$^{2}$.

We also use the HST/NICMOS (Near-Infrared Camera and Multi-Object
Spectrometer) UDF photometry data for our analysis, which were taken
with NICMOS under the HST Cycle 12 Treasury Program in two bands,
F110W and F160W, which roughly correspond to $J$ and $H$. Due to the
small field of NICMOS camera, the NICMOS UDF only covers a part
 (5.76 arcmin$^{2}$) of the optical UDF. The data reduction and 
photometry were introduced by Thompson \etal (2005). 
The 5-$\sigma$ limiting AB magnitude  is 27.7 at 1.1 and 1.6 $\mu$m 
in a 0.6$''$ diameter aperture.

Ground-based near-infrared images ($JHK$) of the UDF were taken as
part of the GOODS survey with ISAAC on the VLT. These data were
obtained as part of the ESO Large Programme LP168.A-0485(A), and
GOODS/ISAAC Data Release Version 1.5 was used for our analysis. This
data release includes 24 fully reduced VLT/ISAAC fields in $J$ and
$K$ bands covering 159.1 and 159.7 arcmin$^{2}$ of the GOODS region,
respectively, and 19 fields in $H$ band covering 126.7 arcmin$^{2}$.
The ISAAC data were reduced using an improved version of the ESO/MVM
image processing pipeline. The 5-$\sigma$ limiting AB magnitudes at
$J$-band, $H$-band and $K$-band are 25.3, 24.8 and 24.4, in a
2.0$''$ diameter aperture.

A catalog of objects whose redshifts are known from spectroscopic
observations is very important to the construction of a photometric 
redshift code.  For this reason, the GOODS multi-object, optical to 
NIR spectroscopy in CDF-S was used for our analysis (Version 2.0, 
24 December 2005), with the FORS2 on the VLT. We obtained 1204 
spectra of 930 individual
targets, providing in total 943 redshift measurements with quality
flag A, B or C (A=solid redshift, B=likely redshift, C=potential
redshift). Of the targets 725 have an assigned redshift with quality 
flag A, B or C. A full description of the survey can be found in 
Vanzella \etal(2005).  
Forty-eight of the targets are located in the UDF, of which 23 have 
quality flag ``A" and 10 have quality flag ``B".

\subsection{Image Resampling}\label{sec:resamp}
All of these data have been released in fully processed form, and no
additional processing is necessary.  However, the images in different
data sets have different scales and sizes, so they must be resampled
in order to put on the same astrometric grid. The resampling was
done with IRAF's $geomap$ and $geotran$ tasks. The NICMOS $H$ image
was used as the base image to which all the other images were
matched. The NICMOS $J$ image need no remapping as it was already on
the same scale. The optical images from the HST/ACS and the
near-infrared images from VLT/ISAAC were remapped. This changed
their scales from 0.03 and 0.15 to 0.09~$''$~pixel$^{-1}$.
The resampling does not introduce any appreciable shifts in either 
the position or flux.

\begin{figure}
\centering
\includegraphics[width=0.95\columnwidth]{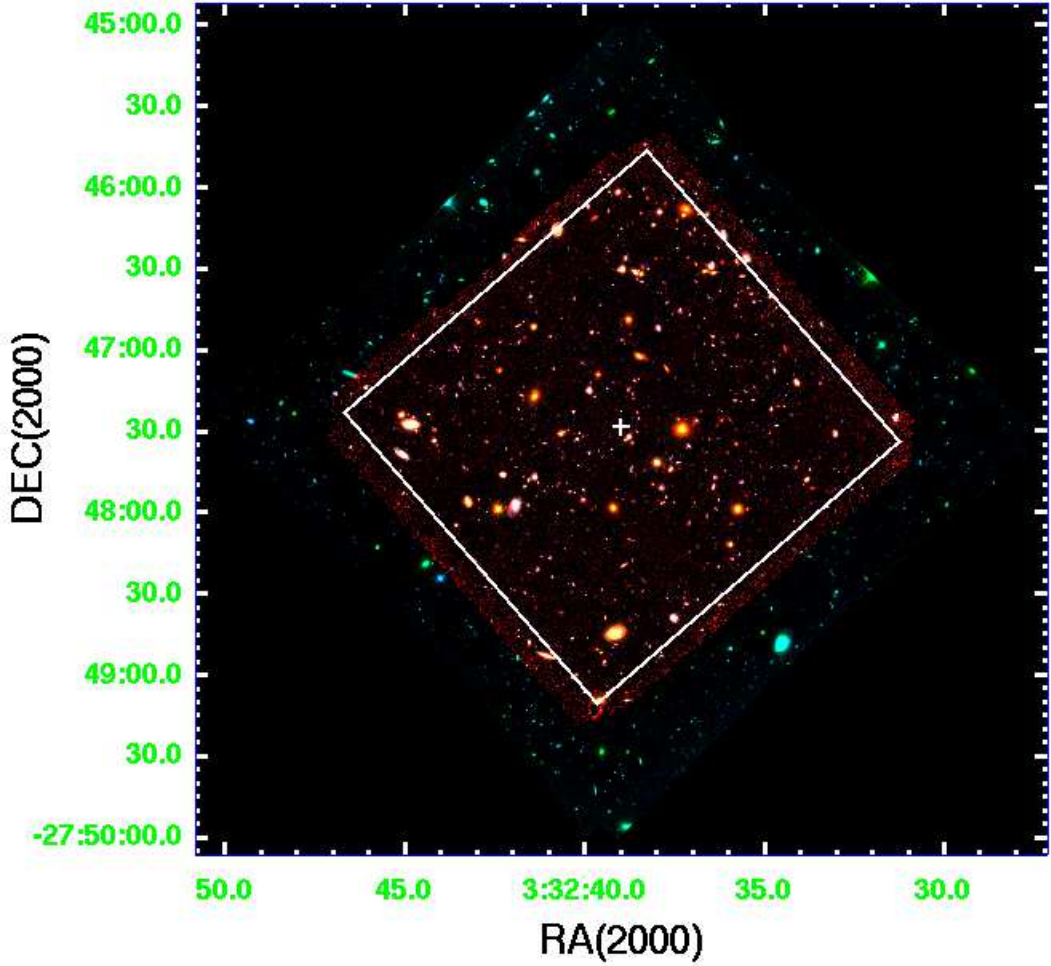}
\caption{
Composite pseudo-color image of the UDF. The RGB colors are assigned
to $H$-,$z$-, and $B$-band images from HST. The outlined (white) 
region near the center of the image is the field where the $JH$-band 
images have a high signal-to-noise ratio (5.5 arcmin$^2$).}
\label{fig:area}
\end{figure}

Figure~\ref{fig:area} shows a composite pseudo-color image of the 
UDF. From this figure, we find that the full drizzled image does not 
have a uniform integration time over the image. In particular, the 
edges of the HST/NICMOS image have only one integration, as compared 
to the average 16 integrations for the interior of the image. 
Therefore, the area, as discussed in this paper, was reduced from 
HST/NICMOS's 5.76 to 5.50 arcmin$^{2}$, as is outlined near the 
center of the image.

\subsection{Photometry}\label{sec:photo}
Source extraction in the science image was performed with the source
extraction program \bsex\ version 2.5 (Bertin \& Arnout 1996) in the
dual image and rms image mode. The source extraction includes four
HST/ACS images, three VLT/ISAAC images, as well as two HST/NICMOS
images. The NICMOS $J+H$ image was used as the detection image for
our catalog. 
The
source extraction parameters were similar to those 
used in the NICMOS treasury version 2.0 catalog, except for 2$''$
diameter aperture used for 
the aperture magnitudes. 
620 sources were 
detected in the
central 5.5\,arcmin$^2$ area down to $H_{\rm AB}=26.0$\,mag. For most 
of the objects, our Kron automatic aperture magnitudes (MAG\_AUTO) for 
$BVizJH$ are close to those in Thompson et al. (2005), 
with differences 
generally less than 0.05\,mag.

The total 
magnitude
was
then defined as the brighter 
of the
Kron automatic aperture 
magnitude
and the corrected aperture 
magnitude. 
The
colors were measured using 
the 2$''$ diameter aperture magnitudes. All magnitudes were corrected for 
Galactic
extinction, $A_B= 0.033$, taken from Schlegel \etal (1998), using
the empirical selective extinction function of Cardelli \etal
(1989). Compared to the optical selection, the near-IR selection (in
particular in the $K$ band) offers several advantages (Cimatti \etal
2002). Therefore, we select objects to $\Kv<22.0$ over a total sky
area of 5.5\,arcmin$^2$ in the UDF, and 210 objects were included in
our final catalog. Simulations of point sources show that in all the
area the completeness is above 90\% at this $K$-band level.

\begin{figure}[h]
  \begin{minipage}[t]{0.50\linewidth}
  \centering
  \includegraphics[width=70mm,height=68mm,angle=-90]{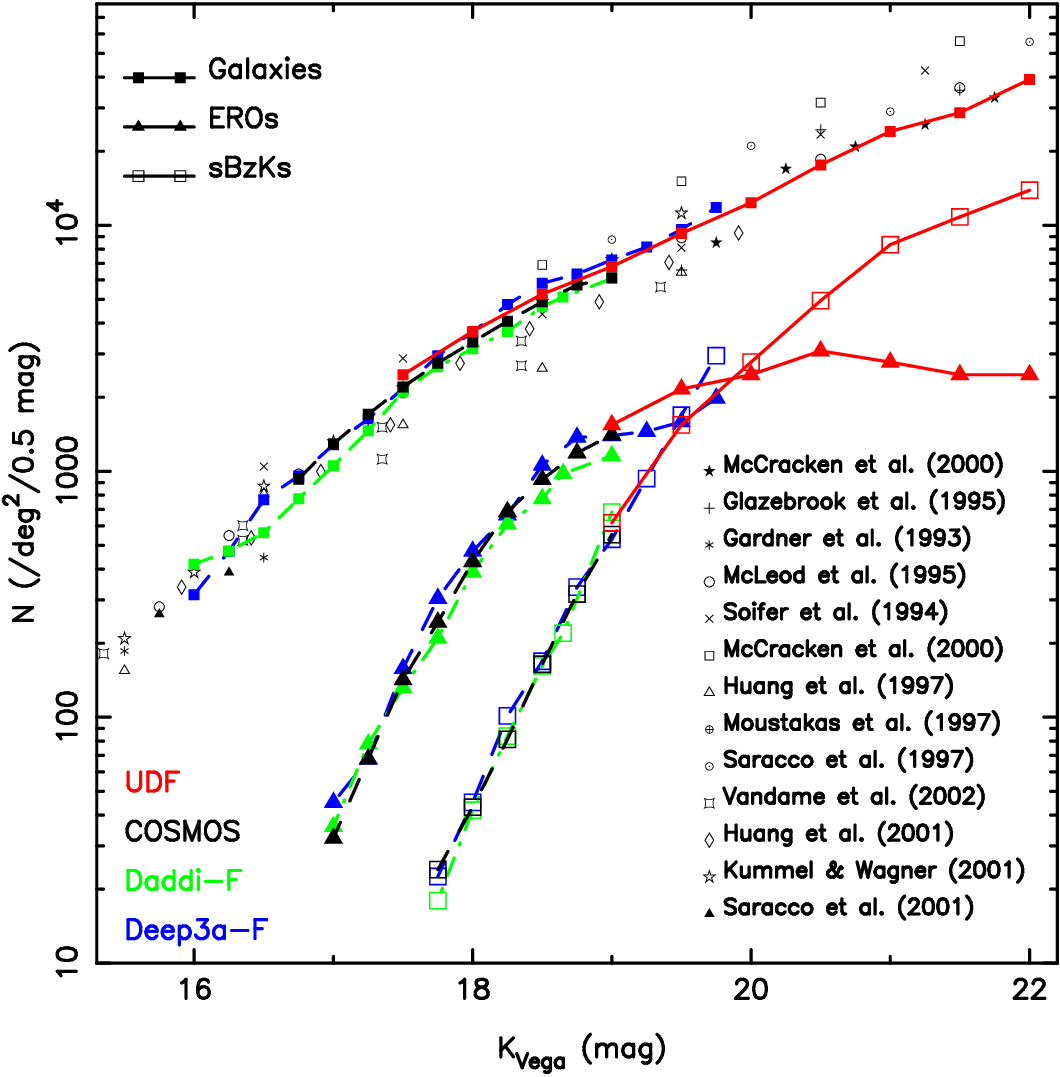}
  \vspace{-5mm}
  \caption{{\small $K$-band differential number counts for field galaxies,
EROs and \sbzks\ (see Sect.~4) in the UDF, compared with a
compilation of results taken from various sources. Filled squares,
triangles and open squares 
represent
the number counts 
of $K$-selected field galaxies,  EROs and \sbzks, respectively.
  } }
  \label{fig:numcz}
  \end{minipage}%
\begin{minipage}[t]{0.50\textwidth}
  \centering
  \includegraphics[height=68mm,angle=-90]{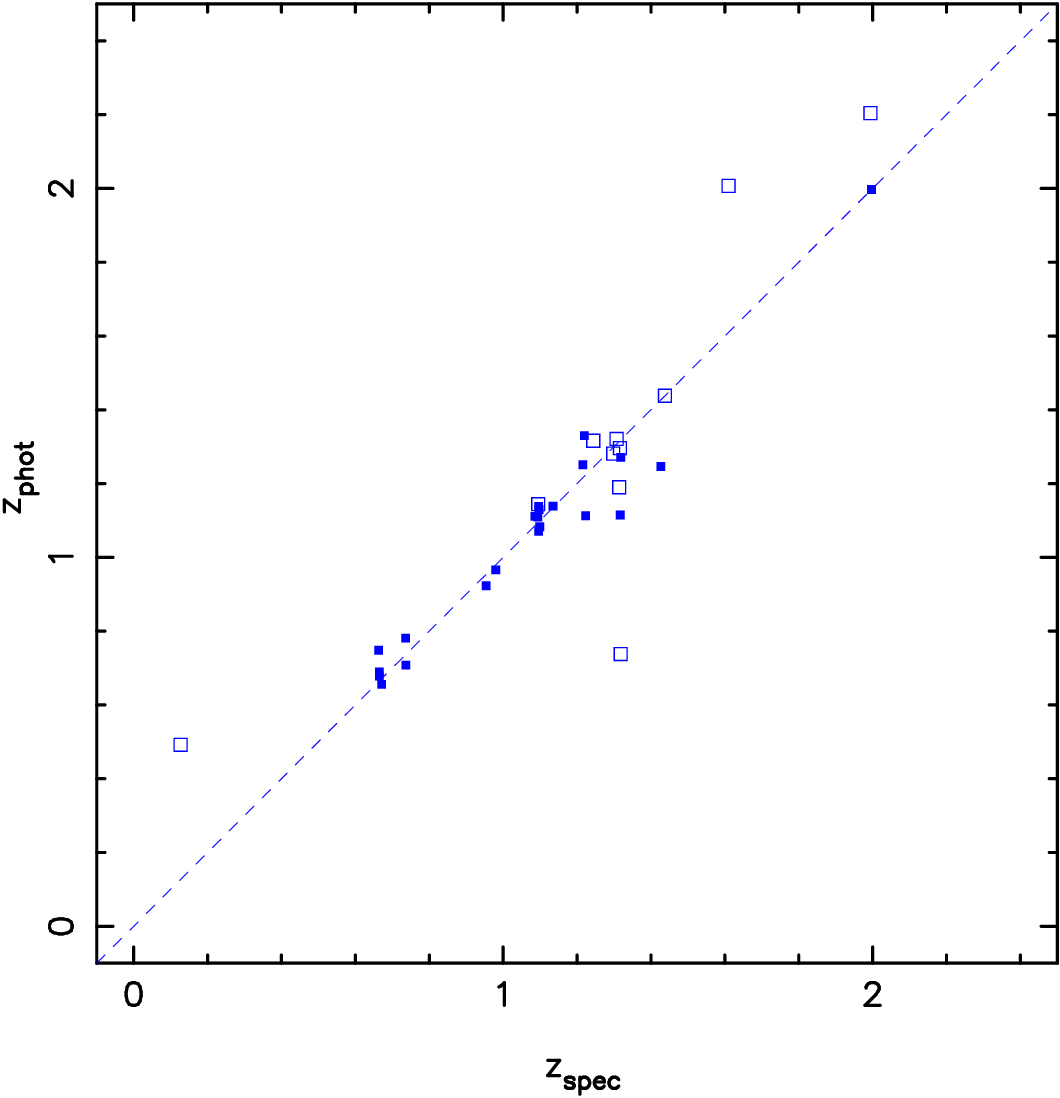}
  \vspace{-5mm}
  \caption{{\small
  Comparison between photometric ($\zph$) and spectroscopic redshifts
($\zsp$) for the UDF spectroscopic sample. Filled squares and open
squares 
represent
respectively 
objects with solid redshifts (quality flag A) and likely redshifts (quality flag B).}}
  \label{fig:photoz}
  \end{minipage}%
\end{figure}

Figure~\ref{fig:numcz} shows a comparison of 
the $K$-band number counts in the UDF survey with a compilation of 
counts
published
in
the
literature. No corrections for incompleteness were applied to our data, and we
excluded stars using the same method as that in Kong \etal (2006).
The red-, black-, green- and blue-filled squares correspond to the
counts of field galaxies in the UDF, COSMOS (Renzini \etal 2008),
Daddi-F and Deep3a-F (Kong \etal 2006), respectively. As shown in
the figure, our number counts in different fields are in good
agreement with those of 
the previous surveys.

\section{Photometric Redshifts}\label{sec:photz}
Photometric redshifts (hereafter $\zph$) were calculated for all
the $\Kv \lsim 22$ objects in the UDF. We take the public code
$hyperz$, which uses 
a usual template-fitting and a $\chi^2$ minimization method 
(Bolzonella \etal 2000).

We use a stellar population synthesis model (Kodama \& Arimoto 1997,
hereafter KA97) to make template SEDs. KA97 was successfully used to
obtain photometric redshifts of low redshift galaxies (Kodama, Bell
\& Bower 1999) and high redshift galaxies (Furusawa \etal 2000). The
template SEDs consisted of the spectra of pure disks, pure bulges
and composites 
obtained by 
interpolation. In total, our basic template set consisted of 303 SEDs (more 
description in 
a
forthcoming paper). The spectral energy distribution derived from
the observed magnitudes of each object was compared to each template
spectrum (redshift from 0.0 to 5.0 
at 
steps
of  0.05; $A_{\rm V}$ from 0.0 to 4.5 
at 
steps
of 0.05 and Calzetti \etal's internal reddening law) in turn. The weighted 
mean redshift from $hyperz$ was 
then calculated for each object.

Figure~\ref{fig:photoz} shows a comparison of the photometric
redshifts with the 33 spectroscopic redshifts from VLT/FOS2
spectroscopy (Sect.~\ref{sec:obs}). Filled squares show the objects
with "solid redshifts" (quality flag A) while open squares show the
objects 
with less
secure 
spectroscopic redshift identifications
(quality flag B, "likely redshift"). Taking into account only the
objects with solid redshifts, the 
error 
in
the
photometric redshifts 
is
$\sigma_z = 0.003 (1+z)$ with no catastrophic failures, but taking
into account all the 33 objects, the error would increase to
$\sigma_z = 0.02 (1+z)$.

\section{$\lowercase{\rm s}$B$\lowercase{\rm z}$K$\lowercase{\rm
s}$ in the UDF}\label{sec:cand}
In this section, we select \sbzks\ in the UDF using the multicolor
catalog based on the NIR $K$-band image. Then we will discuss 
the
number counts, clustering, morphologies, reddening, 
SFR and stellar masses of faint \sbzks\ in the UDF.

\subsection{\sbzks\ Selection}
The UDF $z$- and $K$-band filters are the same as those used in
Daddi \etal (2004) while the $B$-band filters are different in the
UDF (F435W) and Daddi \etal (2004) (Bessel B). Using the Pickles
(1998) stellar spectra, we find the correction term to the $B$-band
is very small ($<0.05$ in all cases). 
So we 
applied the $BzK$ selection criterion from Daddi \etal (2004) 
without
change
to 
the \sbzks\ selection in the UDF without additional correction.

\begin{figure*}
\centering
\includegraphics[angle=-90,width=0.95\columnwidth]{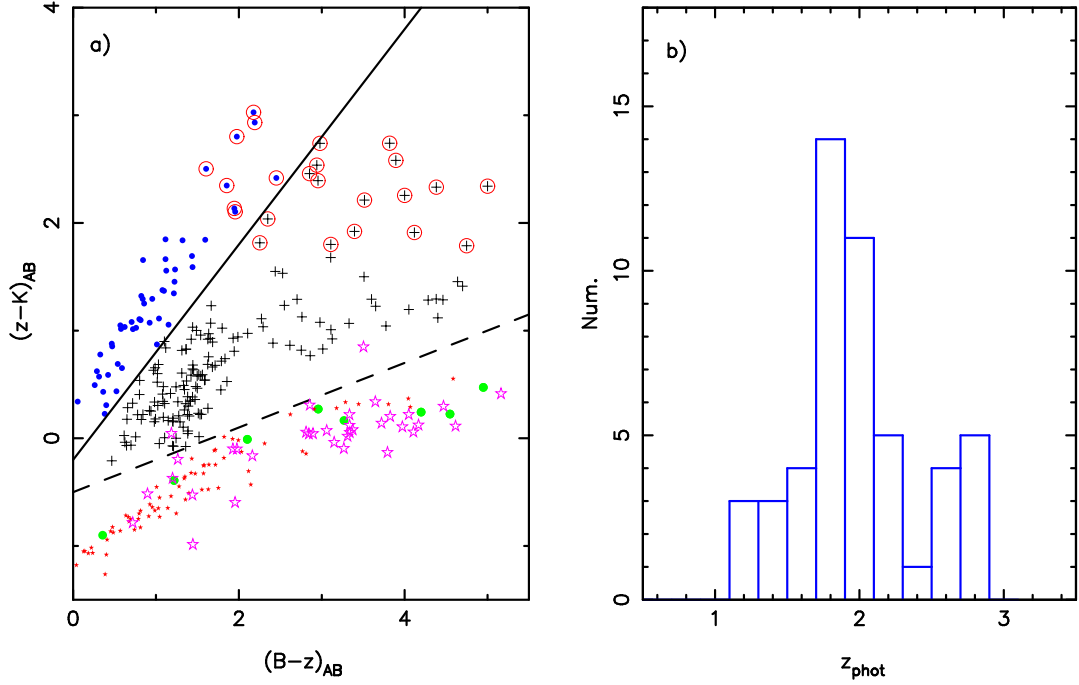}
\caption{
a) The $(z-K)_{\rm AB}$ vs $(B-z)_{\rm AB}$ color-color diagram for
the galaxies in the UDF. The diagonal solid line defines the region
$BzK\equiv (z-K)_{\rm AB}-(B-z)_{\rm AB}\geq-0.2$ that is
efficient 
for 
isolating $z>1.4$ star forming galaxies (\sbzks). The dashed line separates 
the regions occupied by stars and galaxies. Solid circles show objects classified 
as stars, having $(z-K)_{\rm AB} - 0.3(B-z)_{\rm AB} < -0.5$; open stars show 
stellar objects from the K20 survey (Daddi \etal 2004); squares represent \sbzks;
circles represent galaxies with $(i-K)_{\rm AB}>2.40$ (EROs); solid
stars correspond to stellar objects given in Pickles (1998).
b) Photometric redshift histogram of 52 $\Kv \lsim 22$ UDF
galaxies selected with the $BzK$ criterion (\sbzks).
}
\label{fig:bzkccd}
\end{figure*}

Figure~\ref{fig:bzkccd}a shows the $BzK$ color diagram of
$K$-selected objects in the UDF. Using the color criterion from
Daddi \etal (2004), $BzK\equiv (z-K)_{\rm AB}-(B-z)_{\rm AB}>-0.2$,
52 galaxies with $K_{\rm Vega} <22$ are selected in the UDF as
\sbzks, which occupy a narrow range on the left of the solid line in
Figure 4a. To $\Kv<22$, the surface density of \sbzks\ is $\sim
9.45\pm1.05$ arcmin$^{-2}$. We also selected EROs in the UDF with
the criterion $i-K>2.4$ (AB mags), corresponding closely to $R-K>5$
(Vega mags). In the UDF to $\Kv<22$ 24 EROs are selected and plotted
in Figure~4a with open circles.

Figure~\ref{fig:bzkccd}b shows the photometric redshift histogram
of 52 $\Kv \lsim 22$ UDF galaxies selected with the $BzK$
criterion. The average redshift and the peak of the redshift
distribution of \sbzks\ are $\zph \sim 1.8$. Only about 10\% of
them (6 objects) have $\zph<1.4$, three of these six objects are
EROs, and the other three are just marginally inside the $BzK$
selection region. The photometric redshift distribution supports
the
view
that the $BzK$ criterion is valid for 
picking
out star-forming galaxies at 
$1.4\lsim z \lsim2.5$ with $\Kv=22.0$.

\subsection{Morphologies of \sbzks}\label{sec:morp}
\begin{figure*}
\centering
\includegraphics[angle=-90,width=0.95\columnwidth]{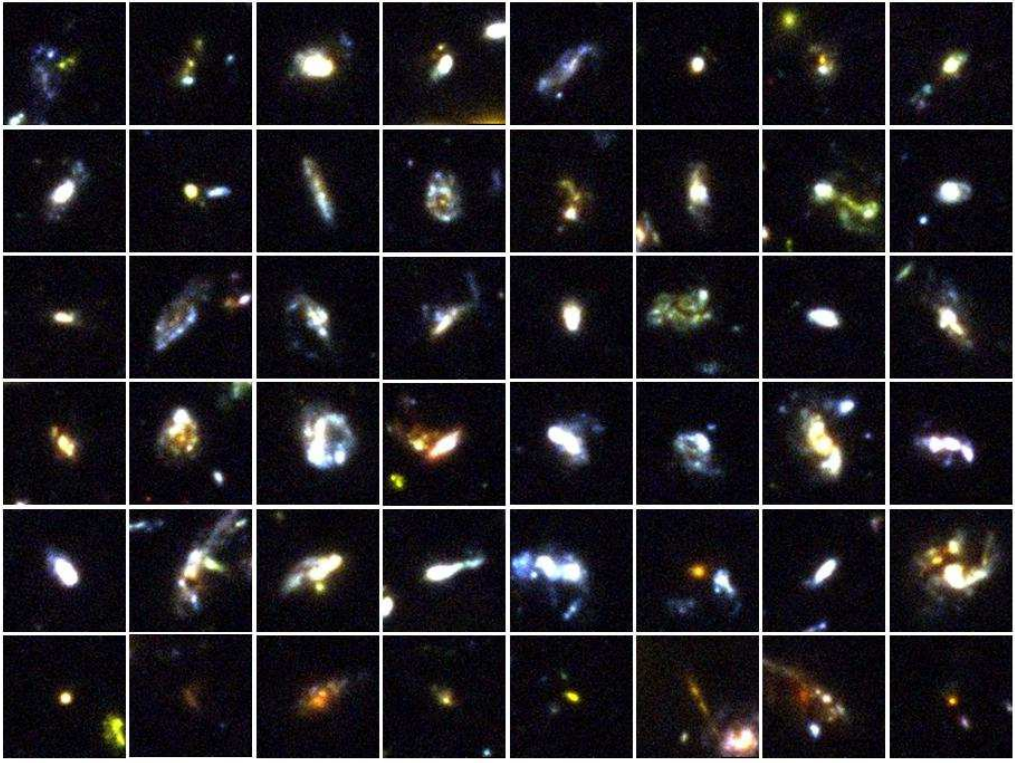}
\caption{
Composite pseudo-color images ($3^{\prime\prime}\times
3^{\prime\prime}$) of 48/52 \sbzks\ in the UDF. The RGB colors are
assigned to $z$-,$i$- and $B$-band images from HST/ACS. Most of
\sbzks\ 
show irregular/merging-like morphologies.
}
\label{fig:morp}
\end{figure*}

The
morphology of a galaxy reflects its dynamical history and evolution.
For example, different Hubble types are associated with different
star formation histories and different 
patterns 
of
motion 
of stars and 
gas. Morphological studies of distant objects have become possible
now with the high-resolution capabilities of HST/ACS. 
The
ACS
imaging
provides a fundamental complement 
of 
investigating the nature of $\sbzks$ and 
elucidating their evolutionary status. Figure~\ref{fig:morp} shows the color images 
for 48 \sbzks\ (to show the image well, we randomly selected 48 \sbzks\ from 
all 52 \sbzks) in the UDF.  
The $B$-, $i$- and $z$-band images were 
colored
red, green, and blue,
respectively.

From this figure, we find that although 
the \sbzks\ show a range of
morphologies from compact protobulges to disturbed 
major-mergers,
most of 
them  
show irregular/merging-like morphologies. The bottom row in Figure~5 shows 
objects which satisfy both the $BzK>-0.2$ criterion (\sbzks) and the $i-K>2.4$ 
criterion (EROs). More 
detailed analysis, including concentration ($C$),
asymmetry ($A$), clumpiness ($S$), 
the Gini coefficient ($G$) and $M_{20}$ for 
the \sbzks\ and comparison with other high redshift
galaxies
will be presented in our forthcoming paper.

\subsection{Number Counts of \sbzks}
The differential number counts 
of \sbzks\ and EROs in the UDF are plotted in Figure~\ref{fig:numcz}. Open 
squares and solid triangles with solid lines in Figure 2 show the number counts 
for 
the \sbzks\ and EROs in the UDF, respectively.

From Figure 2 some 
characteristics are 
distinguished:
1) The fraction of \sbzks\ increases very steeply towards fainter
magnitudes, and the counts of \sbzks\ have roughly the same slope
at all K-band magnitudes. 2) The slope of the number counts of
EROs is variable, being steeper at bright magnitudes and
flattening out towards faint magnitudes. A break in the counts is
present at $\Kv\sim 18.0$, very similar to the break that has been
shown in the previous works. 3) The counts of field galaxies, EROs
and \sbzks\ in the UDF are almost identical to those in COSMOS,
Daddi-F, and Deep3a-F to their limits of $\Kv\sim20$.

\subsection{Clustering of \sbzks} \label{sec:clustering}
Measurement
of
the clustering of galaxies provides an additional tool 
for studying the evolution of galaxies and the formation of
structures. In this section we estimate the angular correlation of
the general galaxy population as well as that of \sbzks. In order
to measure the angular correlation function of various galaxy
samples, we apply the Landy-Szalay technique (Landy \& Szalay
1993; Kerscher \etal 2000). A fixed slope $\delta=$\ 0.8 is
assumed for the two-point correlation function.

\begin{figure*}
\centering
\includegraphics[angle=-90,width=0.95\columnwidth]{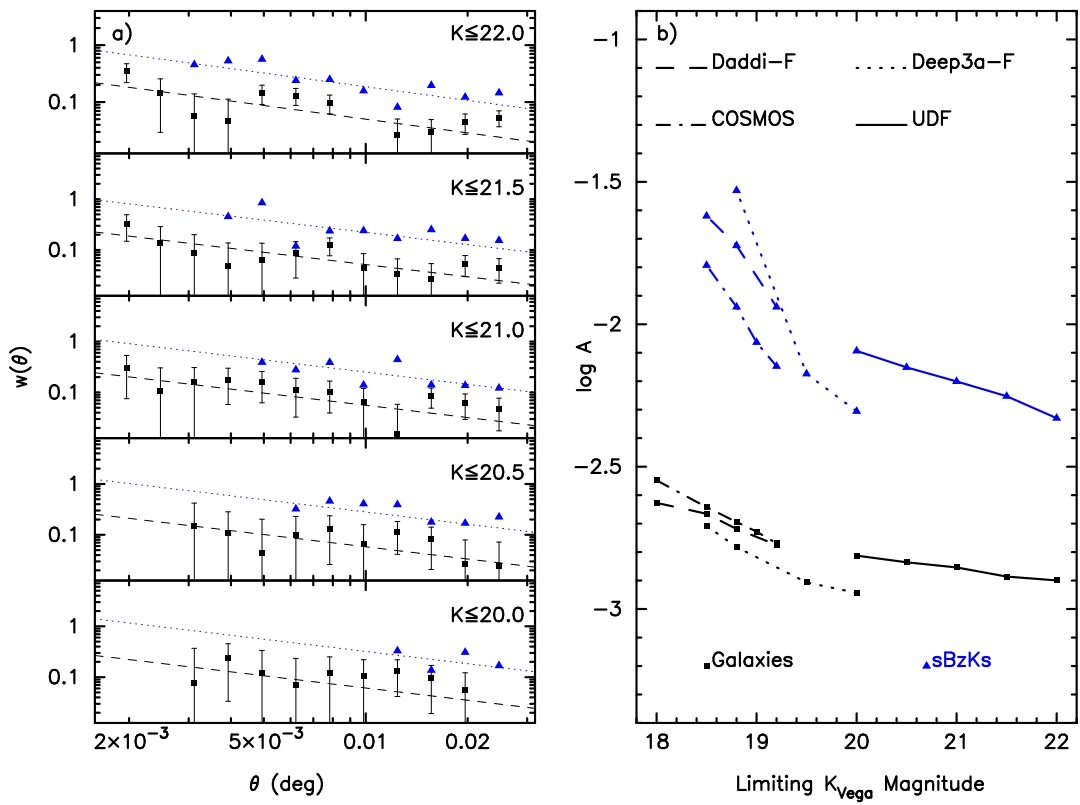}
\caption{ a)Observed, bias-corrected two-point correlations of
field galaxies (squares) and \sbzks\ (triangles). The error bars
on the direct 
estimated values are $1 \sigma$ errors. The lines
show the power-law fitted to the $w(\theta)$. b) Angular
clustering amplitudes of field galaxies and \sbzks\ in different
fields are shown as a function of $K$-band limiting magnitudes. }
\label{fig:clus}
\end{figure*}

In Figure~\ref{fig:clus}a, the bias-corrected two-point
correlation functions $w(\theta)$ of field galaxies are shown as
squares. The dashed line shows the power-law correlation function
given by a least squares fitting. 
We clearly 
have detect a positive correlation 
for 
the field galaxies
with an angular dependence broadly consistent with the adopted slope
$\delta=$\ 0.8. The derived clustering amplitudes (where $A$ is the
amplitude of the true angular correlation at 1 degree) are $A=1.54$,
1.46, 1.40, 1.30 and 1.26 for $\Kv<20.0$, 20.5, 21.0, 21.5, 22.0,
respectively.

We estimate the clustering properties of the UDF \sbzks\ 
(the triangles
in Figure 6a). The
dotted lines show the power-law correlation function given by a
least squares fitting to the measured values. A strong clustering
of \sbzks\ is indeed present at all scales,
with 
amplitudes 
about 6 times 
higher than
those of the field population at the same $K_{\rm Vega}$ limits,
in agreement with 
the previous findings (Kong et al. 2006). The derived 
amplitudes are $A=8.07$, 7.06, 6.30, 5.59 and 4.68 for $\Kv<20.0$,
20.5, 21.0, 21.5, 22.0, respectively.

Figure~6b summarizes the clustering measurements for the 
two
populations examined (field galaxies and \sbzks), as a function of the $K$-band
limiting magnitudes of the samples. Clear trends with 
the $K$ magnitudes are present for all samples, that 
the
fainter galaxies 
apparently
have 
lower
intrinsic (real space) clustering, consistent with the fact that
objects with fainter $K$ magnitudes are less massive, or have wider
redshift distributions, or both. We compare our results on the UDF
with those previously reported by Renzini \etal (2008) for the
COSMOS and Kong \etal (2006) for the Daddi-F and Deep3a-F, and find
they all show a smooth decline in amplitude with $K$-band
magnitudes. However, the decline is not 
less
steep 
in the
range $18<\Kv<20$. Furthermore, the clustering amplitudes in these
fields are slightly different. The field-to-field variation may be
one of the possible reasons for this discrepancy, depending on the
survey geometry, surface density and clustering properties.

\subsection{Physical Quantities of \sbzks}\label{sec:phy}
In this subsection,
physical properties of \sbzks\, such as color excess $E(B-V)$, SFRs
and stellar masses, are derived on the basis of the present
photometric data and the recipes calibrated in Daddi \etal (2004).

\begin{figure*}
\centering
\includegraphics[height=7.0cm,width=0.55\columnwidth]{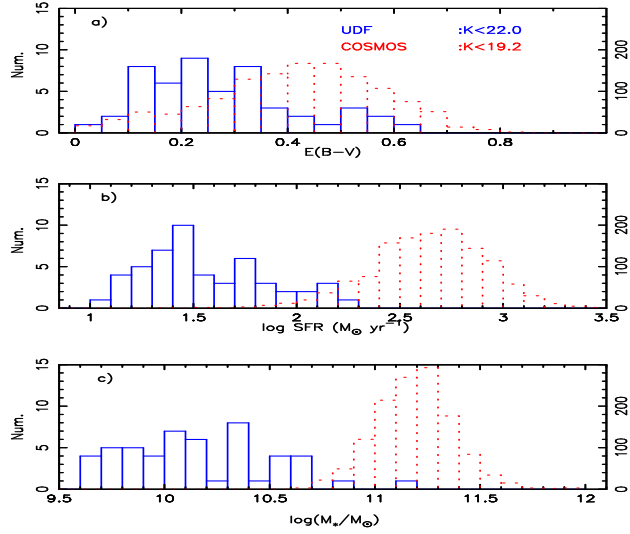}
\caption{
Panels (a) - (c) show, respectively,
the histograms of 
reddening, star formation rate 
and stellar mass.
Solid 
histograms 
refer
to \sbzks\ in the UDF (numeric label
on the left),
dotted histograms,
to the COSMOS 
objects
(numeric labels 
on the right).
}
\label{fig:bzk}
\end{figure*}

Following Daddi \etal (2004), 
the reddening of \sbzks\ can be
estimated by the $B-z$ colors, providing a measure of the UV
slope. The Daddi \etal's recipe is consistent with the recipes by
Meurer \etal (1999) and Kong \etal (2004) for the Calzetti \etal
(2000) extinction law, based on the UV continuum slope. The
$E(B-V)$ histogram of \sbzks\ in the UDF is shown in
Figure~\ref{fig:bzk}a (solid line). The median reddening for the
$\Kv<22$ \sbzks\ is estimated to be $E(B-V)=0.28$, and 18\% of
\sbzks\ have $E(B-V)>0.4$, where the UV-based criterion of Steidel
\etal (2004) would fail to select the galaxies at $z\sim2$. For
comparison, we show the $E(B-V)$ histogram of \sbzks\ in the
COSMOS ($\Kv<19.2$) in Figure~\ref{fig:bzk}a (dotted line), too.
The median reddening for the $\Kv<19.2$ \sbzks\ in the COSMOS is
estimated to be $E(B-V)=0.42$, and 56\% of them have $E(B-V)>0.4$.
The difference is due to the different limiting $K$ magnitudes in
the two fields.

Knowing
the reddening, the reddening corrected $B$-band flux
is used to estimate the 1500\,\AA\ rest-frame luminosity, assuming
an average redshift of 1.8, which can 
then
be translated into SFR on the 
basis of the Bruzual \& Charlot (2003) models. Daddi et al. (2005)
showed that SFRs derived in this way were consistent with radio and
far-IR based estimates for the average \sbzks. The SFR histograms of
\sbzks\ in the UDF and COSMOS are shown in Figure 7b as solid line
and dotted line, respectively. About 16\% of the \sbzks\ in the UDF
have SFR$>70\,M_\odot$~yr$^{-1}$, and the median SFR is about
$36~M_\odot$~yr$^{-1}$. However, about 98\% of the \sbzks\ in the
COSMOS have SFR$>$70$\,M_\odot$~yr$^{-1}$, and the median SFR is
$430~M_\odot$~yr$^{-1}$.

Following Daddi \etal (2004), 
the stellar masses of 
the \sbzks\ are estimated 
by an empirical relation 
between their observed $K$-band total magnitudes and $z-K$ colors. The 
histograms for the stellar masses of \sbzks\ derived in this way are shown in 
Figure 7c. Almost all of the \sbzks\ in the UDF have $M_*<10^{11} M_\odot$
and the median stellar mass is $1.4\times10^{10}M_\odot$; in the
COSMOS, $\sim 85\%$ of the \sbzks\ have $M_*>10^{11} M_\odot$ and
the median stellar mass is $\sim 1.5\times10^{11}M_\odot$. The
higher masses for 
the \sbzks\ in the COSMOS compared to the UDF result
from the shallower $K$-band limit.

\begin{figure*}
\centering
\includegraphics[width=0.5\columnwidth]{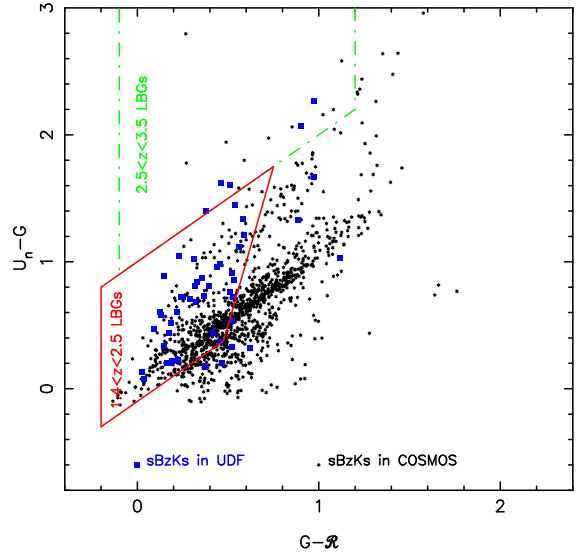}
\caption{ $U_{\rm n}GR$ color-color diagram for \sbzks\ in the UDF
  (squares) and COSMOS (stars). The $U_{\rm n}GR$ colors were derived
  from 
SED fitting. The solid line shows color regions defined
  for the UV identification at $z\sim2$ and $z\sim3$.
}
\label{fig:ugr}
\end{figure*}

\section{\sbzks- vs. UV-selected Galaxies at $z\sim2$}
\label{sec:UV}
Recently, the UV technique for selecting LBGs has been extended to
$z<3$ using a $U_{\rm n}GR$ color-color diagram, which isolates
star-forming galaxies at $1.4<z<2.5$ (Erb \etal 2003; Steidel \etal
2004; Adelberger \etal 2004), a redshift range matching that of the
$BzK$ selection. To compare these different color criteria
(BzK-criterion and UV-criterion) for selecting star-forming galaxies
at $1.4<z<2.5$, we show the \sbzks\ in the UDF and COSMOS on
the $U_{\rm n}GR$ color-color diagram in Figure~\ref{fig:ugr}.

As no $U_{\rm n}GR$ photometry is available to us for the \sbzks\
in the UDF region, synthetic $U_{\rm n}GR$ magnitudes have been
derived from the KA97 models 
which
provided
the best fits to the observed ACS-$BViz$, NIC-$JH$ and ISAAC-$JHK$ SEDs. 
Figure 8 shows the resulting synthetic $(G-R)$ vs. $(U_n-G)$ colors. Among 
the 52 $\Kv<22$ \sbzks\ in the UDF, 39 (75\%) would be selected by the
Adelberger \etal (2004) criterion for the galaxies at $z\sim2$,
five would be selected by the Steidel \etal (2003) criterion for
the galaxies at $z\sim3$, only eight $\Kv<22$ \sbzks\ fail to be
selected by the UV criterion 
(these are highly dust obscured galaxies).

In Figure~\ref{fig:ugr}, we also plot the COSMOS \sbzks\ on the
synthetic $(G-R)$ vs. $(U_n-G)$ diagram. Synthetic
$U_{\rm n}GR$ magnitudes have been derived from the KA97 models
that
provided
the best fits to the observed CFHT-$ui$, Subaru-$BVg'r'i'z'$, SDSS-$u'g'r'i'z'$ 
and KPNO/CTIO-$Ks$ SEDs. Among $\sim1300$ $K<19.2$ \sbzks\ in the 
COSMOS, only 450 (35\%) would be selected by the Adelberger \etal (2004) 
criterion while about 65\% $\Kv<19.2$ \sbzks\ fail to be selected by the UV criterion.

From Figure~8 we find that most of the faint \sbzks\ can be selected
by the UV criterion for the galaxies at $z\sim2$, but it does not
work well for the 
brighter \sbzks\ 
objects. One possible reason is that the UV-selection requires the UV continuum 
to be relatively flat, thus limits the overall dust extinction to $E(B-V)\lsim0.4$
(Adelberger \& Steidel 2000), but many of the bright \sbzks\ at
$z>1.4$ are more reddened. As we discussed in Section~4.5, a large
proportion of highly reddened galaxies 
are actively star-forming and
have
large
stellar masses. Therefore, a large fraction (as high as
$\sim 65\%$) of the $K<19.2$ galaxies at $z>1.4$ 
fail
to be selected by the UV criterion 
of 
identifying galaxies at $z\sim2$. As a consequence, a significant 
part of the SFR density and stellar mass density 
is missed.

\section{Summary and Conclusions}\label{sec:summary}
In this paper, we investigate the properties of star-forming
galaxies at $1.4 \lsim z \lsim 2.5$ (\sbzks) in the HST Ultra Deep
Field. Using images from ACS-$BViz$, NIC-$JH$, ISAAC-$JHK$ and
spectroscopy from VLT/FOS2, we analyze the photometric
redshifts, number counts, clustering, $E(B-V)$, SFRs, stellar masses
and morphologies of \sbzks\ in the UDF. The main results can be
summarized as follows.

1. Based on 
the KA97 templates and a standard $\chi^2$ minimizing
method, we have developed a new photometric redshift method.
Comparing the photometric redshifts for 33 galaxies with
spectroscopic redshifts from VLT/FOS2, we find that the 
error
in
the
photometric
redshifts
is 
less than 0.02(1+$z$).

2. Combining multiwave band ground-based and ACS/NIC photometric
data, we estimate 
the photometric redshifts for galaxies in the UDF. We find the $BzK$ color 
criterion can be used to select star-forming galaxies at $1.4 \lsim z \lsim 2.5$ 
with $\Kv<22.0$.

3. Down to $\Kv<22.0$, we 
selected 52 \sbzks\ in the UDF. Most of the \sbzks\ have irregular morphologies 
on the ACS images. The 
logarithm of the number counts of \sbzks\ increases linearly with the $K$ magnitudes,
while that of EROs flattens out by $\Kv \sim 19$.

4. 
The
two-point angular correlation function of \sbzks\ and field
galaxies 
were calculated, and the clustering amplitudes of \sbzks\
are 
found
to
be about a factor of 6 higher than those of generic galaxies 
at the same magnitude range. A smooth decline in amplitude with $K$-band
magnitude
is found, which is consistent with results from previous works.

5. Using 
the approximate relations from Daddi et al. (2004) and
multicolor photometry, we 
estimated the color excess, SFRs and stellar masses of \sbzks. These $\Kv<22$ 
galaxies have a median reddening of $E(B-V)\sim0.28$, an average SFR of $\sim36\
M_{\odot}$~yr$^{-1}$, and a typical stellar mass of $\sim10^{10}M_\odot$.

6. The UV criterion for the galaxies at $z\sim2$ can select most of
\sbzks\ in the UDF ($\Kv\lsim22.0$), but it does not work well for the
brighter 
\sbzks\ sample in the COSMOS ($\Kv\lsim19.2$). A large fraction
of massive, highly reddened, actively star-forming galaxies can not be
selected by the UV criterion.

\begin{acknowledgements}
We are grateful to Prof. Qirong Yuan for valuable suggestions. The
work is supported by the National Natural Science Foundation of
China (NSFC, Nos. 10573014 and 10633020), the Knowledge Innovation
Program of the Chinese Academy of Sciences (No. KJCX2-YW-T05), and
National Basic Research Program of China (973 Program) (No.
2007CB815404). Observations have been carried out using Hubble Space
Telescope (HST) and the Very Large Telescope (VLT).
\end{acknowledgements}

\label{lastpage}
\end{document}